\documentclass[
superscriptaddress,
amsmath,
amssymb,
aps,
pra,
twocolumn,
floatfix,
longbibliography
]{revtex4-1}

\usepackage{graphicx}
\usepackage{dcolumn}
\usepackage{bm}
\usepackage{subfigure}
\usepackage[usenames,dvipsnames]{xcolor}
\usepackage{physics}
\usepackage{mathrsfs}
\usepackage{units}
\usepackage[colorlinks,urlcolor=blue,citecolor=blue,linkcolor=blue]{hyperref}
\usepackage{upgreek}

\begin{document}

\title{Wide-bandwidth atomic magnetometry via instantaneous-phase retrieval}

\author{Nathanial Wilson}
\author{Christopher Perrella}
 \email{chris.perrella@adelaide.edu.au}
 \affiliation{
 Institute for Photonics and Advanced Sensing (IPAS), and School of Physical Sciences\\
 The University of Adelaide, South Australia 5005, Australia
}
\author{Russell Anderson}
 \affiliation{
 La Trobe Institute for Molecular Science, and School of Molecular Sciences\\
 La Trobe University, Victoria 3552, Australia
}
\author{Andr\'{e} Luiten}
 \email{andre.luiten@adelaide.edu.au}
\author{Philip Light}
 \affiliation{
 Institute for Photonics and Advanced Sensing (IPAS), and School of Physical Sciences\\
 The University of Adelaide, South Australia 5005, Australia
}

\begin{abstract}
We develop and demonstrate a new protocol that allows sensing of magnetic fields in an extra-ordinary regime for atomic magnetometry. Until now, the demonstrated bandwidth for atomic magnetometry has been constrained to be slower than the natural precession of atomic spins in a magnetic field---the Larmor frequency. We demonstrate a new approach that tracks the instantaneous phase of atomic spins to measure arbitrarily modulated magnetic fields with frequencies up to fifty times higher than the Larmor frequency.
By accessing this regime, we demonstrate magnetic-field measurements across four decades in frequency up to \unit[400]{kHz}, over three orders of magnitude wider than conventional atomic magnetometers. Furthermore, we demonstrate that our protocol can linearly detect transient fields 100--fold higher in amplitude than conventional methods. We highlight the bandwidth and dynamic range of the technique by measuring a magnetic field with a broad and dynamical spectrum.
\end{abstract}

\date{\today}

\maketitle


\section{Introduction}
\label{sec:Introduction}

Measuring magnetic fields with high accuracy and precision is paramount in myriad applications including medical diagnostics and imaging~\cite{bison2009room, bison2003laser, knappe2010cross, johnson2010magnetoencephalography, xu2006magnetic, boto2018moving, jensen2018magnetocardiography, belfi2007cesium, bison2003dynamical, xia2006magnetoencephalography, alem2015fetal}, geomagnetism~\cite{dang2010ultrahigh}, and fundamental physics~\cite{smiciklas2011new, berglund1995new, pustelny2013global, kimball2013dual}.
Superconducting quantum interference devices (SQUIDs)~\cite{schmelz2011field, schonau2013squid} and optical atomic magnetometers (OAMs)~\cite{kominis2003subfemtotesla, dang2010ultrahigh, sheng2013subfemtotesla} reign as exquisitely sensitive detectors of static and slowly changing magnetic fields with $\unit{fT/\sqrt{Hz}}$ precision. However, there are many applications e.g. bio-magnetic signals~\cite{barry2016optical}, magnetic communications~\cite{gerginov2017prospects}, and improvised threat detection~\footnote{UK Defence Security Accelerator, \href{https://gov.uk/government/publications/countering-drones-finding-and-neutralising-small-uas-threats}{\emph{Countering Drones -- Finding and neutralising small UAS threats,}} April 2019.} where there is a desire to detect time-varying magnetic fields.

One pathway to high-speed magnetometry are so-called `resonantly tuned' magnetometers, whose Larmor frequency is proximate to an oscillating magnetic field of interest~\cite{savukov2005tunable,ledbetter2007detection,chalupczak2012room, lee2006subfemtotesla, savukov2014ultra, deans2018subpicotesla}.
These devices are suitable when the signal is relatively narrowband, about a frequency that is known \textit{a priori}, and provided that the dc field strength can be tuned accordingly.
The bandwidth of these magnetometers is set by the spin resonance linewidth $\Gamma \propto 1/T_2$, of order 3 Hz to 400\,Hz~\cite{savukov2005tunable}.
This approach to ac magnetic sensing surrenders two key benefits of dc atomic magnetometry: (a) the output signal is an indirect measure of the oscillating field amplitude, and is no longer calibration-free~\cite{rajroop2018radiofrequency}, while (b) the sensor only responds linearly to magnetic field fluctuations that have a magnitude much less than the resonance width~\cite{ledbetter2007detection}---typically of order $\unit[1]{nT}$ and less.
Although it is possible to enhance the amplitude range and bandwidth of this type of magnetometer through intentionally decreasing the effective spin relaxation time, this necessarily comes at the expense of sensitivity~\cite{gawlik2009new, gawlik2017nonlinear, budker2007optical, budker2013optical, jimenez2012high, pustelny2006influence}.

As an alternative, quantum metrology protocols such as dynamical decoupling~\cite{chaudhry2014utilizing,baumgart2016ultrasensitive, anderson2018continuously}, compressive sensing~\cite{cooper2014timeresolved, puentes2014efficient} or Hamiltonian estimation~\cite{declercq2016estimation} can provide retrospective insight into magnetic waveforms. 
The recent demonstrations of quantum lock-in detection measure the frequency of continuously oscillating fields with superb submillihertz precision~\cite{boss2017quantum, schmitt2017submillihertz,glenn2018highresolution}.
Contemporary approaches have used entanglement to enhance rf field detection~\cite{ciurana2017entanglement}, and predictive filters to track time-dependent signals~\cite{martinez2018signal}.
However, the challenge with these protocols is the need to have prior information about the waveform, or a requirement that its spectrum be single frequency.

None of these aforementioned approaches fulfills an urgent need for real-time and accurate detection of broadband magnetic signals with a mixture of frequencies and amplitudes. We present a new protocol for time-dependent magnetometry that retrieves the instantaneous spin-precession frequency. This permits the observation of magnetic fields that are varying much faster than the spin-precession frequency itself. This new regime of supra-Larmor-frequency modulation has been hitherto unexplored, and perhaps surprisingly, lies outside the realm of conventional FM signal processing.

Our approach exploits the technique of free-induction decay (FID) in a fundamentally original way. The standard approach is to (a) observe the FID of an ensemble of spins~\cite{jasperse2017continuous, bison2018sensitive, hunter2018waveform, afach2015highly, grujic2015sensitive, hunter2018free}, (b) sinusoidal regression to the observed signal to obtain a single estimate of the Larmor frequency, and then (c) repeat this process in a train of optical pumping and free-induction decay cycles~\cite{hunter2018waveform}. This naturally imposes a maximum bandwidth on the order of the reciprocal of the repetition rate, e.g. an FID train with a repetition rate of \unit[1]{kHz} could track magnetic fields oscillating orthogonal to the dc field up to \unit[100]{Hz}~\cite{miao2019wide}, and parallel to the dc field with a bandwidth up to \unit[1]{kHz}~\cite{hunter2018waveform}.

We instead use an innovative protocol to measure the instantaneous phase of precessing spins, rather than their average frequency. In this way we obtain an instantaneous measure of the time-dependent Larmor frequency, in direct proportion to the time-dependent magnetic-field strength. Our approach can thus deliver the key benefits of a low frequency atomic magnetometer, i.e. calibration-free and linear~\cite{kominis2003subfemtotesla, dang2010ultrahigh, sheng2013subfemtotesla, wilson2019ultrastable}.
Moreover, as spin precession has no `inertia'~\cite{budker2007optical, gawlik2009new}, the Larmor frequency will respond instantaneously to a change in the external magnetic field. Here we demonstrate this by way of measurement of single-tone modulations up to 50 times higher than the Larmor frequency, and by measuring arbitrarily modulated magnetic signals with significant spectral content above the Larmor frequency. We note that, to our knowledge, no-one has yet demonstrated a protocol that has allowed access to this extraordinary regime.


\section{Instantaneous phase retrieval}
\label{sec:InstantaneousPhase}

We induce transverse magnetization in an atomic vapor by amplitude modulating an optical pumping beam near twice the Larmor frequency~\cite{alexandrov2005dynamic}.
Upon extinction of the pump beam, Faraday polarimetry of an off-resonant optical probe constitutes a weak measurement of the freely precessing atomic spins~\cite{budker2007optical}.
\footnote{
  Although we establish an atomic alignment~\cite{budker2000sensitive} and subsequently track its precession, the technique described here would be equally applicable to extracting the instantaneous phase of an atomic orientation (oscillating about an $f_{\rm c} = f_{\rm L}$ carrier). We note that the use of an amplitude-modulated pump beam while establishing coherence and the relative orientation of the pump and probe beams are not critical elements of the technique reported here.}
For a dc magnetic-field strength $B_{\rm dc}$, the polarization rotation of the probe beam $\phi\left(t\right) \propto \sin\left(2\pi f_{\rm c}t\right)$ has a `carrier' frequency $f_{\rm c} = 2 f_{\text{L}}$, where $2 \pi f_{\text{L}} = \gamma B_{\text{dc}}$ is the Larmor frequency and $\gamma$ is the gyromagnetic ratio of the ground-state hyperfine level.
In our configuration the polarization-rotation oscillates at twice the Larmor frequency, owing to $\abs{\Delta m_{F}} = 2$ ground-state coherences~\cite{wilson2018simultaneous}.
For a time-varying magnetic field, we will see a polarization rotation:
\begin{equation} \label{OptRotGeneral}
	\phi\left(t\right) = a_{\text{c}}\cos\left(2\pi\int_{0}^{t}f_{\text{I}}\left(\tau\right)\text{d}\tau + \varphi_{\text{c}}\right) \, ,
\end{equation}
where $a_{\text{c}}$ is the carrier amplitude, and $\varphi_{\text{c}}$ is the arbitrary phase of the carrier.
The instantaneous frequency $f_{\text{I}}\left(t\right)$ provides a direct measure of the instantaneous magnetic-field strength $B(t)$ via:
\begin{equation} \label{InstantaneousField}
	B(t) = \frac{\pi f_{\text{I}}(t)}{\gamma} \, .
\end{equation}

To calculate the instantaneous frequency, we retrieve the instantaneous phase of the measured polarization rotation using the analytic representation $\phi_{\text{a}}\left(t\right) = \phi\left(t\right) + i\mathscr{H}\left\{\phi\left(t\right)\right\}$, where $\mathscr{H}$ is the Hilbert transform~\cite{boashash1992estimating}
\begin{equation}
	\mathscr{H}\left\{\phi\left(t\right)\right\} = \frac{1}{\pi}\text{p.v.}\int_{-\infty}^{\infty}\frac{\phi\left(\tau\right)}{t - \tau}\text{d}\tau \, .
\end{equation}
In the above, $\text{p.v.}$ denotes the Cauchy principal value, and the Hilbert transform imparts a $90^{\circ}$ phase shift to every Fourier component of $\phi\left(t\right)$.
The instantaneous phase of $\phi\left(t\right)$ is then estimated via:
\begin{equation} \label{InstantaneousPhaseArg}
	\varphi_{\text{I}}(t) = \text{arg}\left\{\phi_{\text{a}}\left(t\right)\right\} \, .
\end{equation}
Upon unwrapping $\varphi_{\rm I}(t)$ we obtain a continuous function of $t$ and compute the instantaneous frequency via~\cite{barnes1992calculation}:
\begin{equation} \label{InstantaneousFrequency}
	f_{\text{I}}(t) = \frac{1}{2\pi}\frac{\text{d}\varphi_{\text{I}}\left(t\right)}{\text{d}t} \, .
\end{equation}

When the magnetic field is sinusoidally modulated, $B(t) = B_{\rm dc} + \Delta B \cos \left(2\pi f_{\text{m}}t + \varphi_{\text{m}} \right)$, and the instantaneous frequency is given by $f_{\text{I}}\left(t\right) = f_{\text{c}} + \Delta f\cos\left(2\pi f_{\text{m}}t + \varphi_{\text{m}}\right)$, where $\Delta f = \gamma \Delta B / \pi$ is the frequency deviation, $f_{\text{m}}$ is the modulation frequency, and $\varphi_{\text{m}}$ is the arbitrary modulation phase offset. The polarization rotation is then:
\begin{equation} \label{OptRotSinusoidalModulation}
	\phi(t) = a_{\text{c}}\cos\left(2\pi f_{\text{c}}t + \beta\sin\left[2\pi f_{\text{m}}t + \varphi_{\text{m}}\right] + \varphi_{\text{c}}\right) \, ,
\end{equation}
where $\beta = \Delta f/f_{\text{m}}$ is the modulation index. Eq.~\eqref{OptRotSinusoidalModulation} makes plain the means to obtain the parameters of an applied magnetic field from the observed frequency-modulated polarization rotation.
For single-tone modulation, the oscillating field amplitude can be determined directly from the oscillation amplitude of either the instantaneous-frequency ($\Delta B = \pi \Delta f / \gamma$) or the instantaneous phase ($\Delta B = \pi \beta f_{\rm m} / \gamma$).
This can be extended to more general time-dependent magnetic fields that admit a Fourier series, by multiplying the modulation index of each Fourier component by its corresponding frequency (see Sec.~\ref{sec:NonTrivialModulation}).

The approach we have described above is one that appears to be unorthodox in signal processing since we have not imposed any restriction on the modulation frequency being below the carrier frequency. Below we show that we can successfully extract magnetic field modulation signals at higher frequency than the carrier.

\section{Experimental setup}
\label{sec:ExperimentalSetup}

\begin{figure}[t]
\includegraphics[width=\columnwidth]{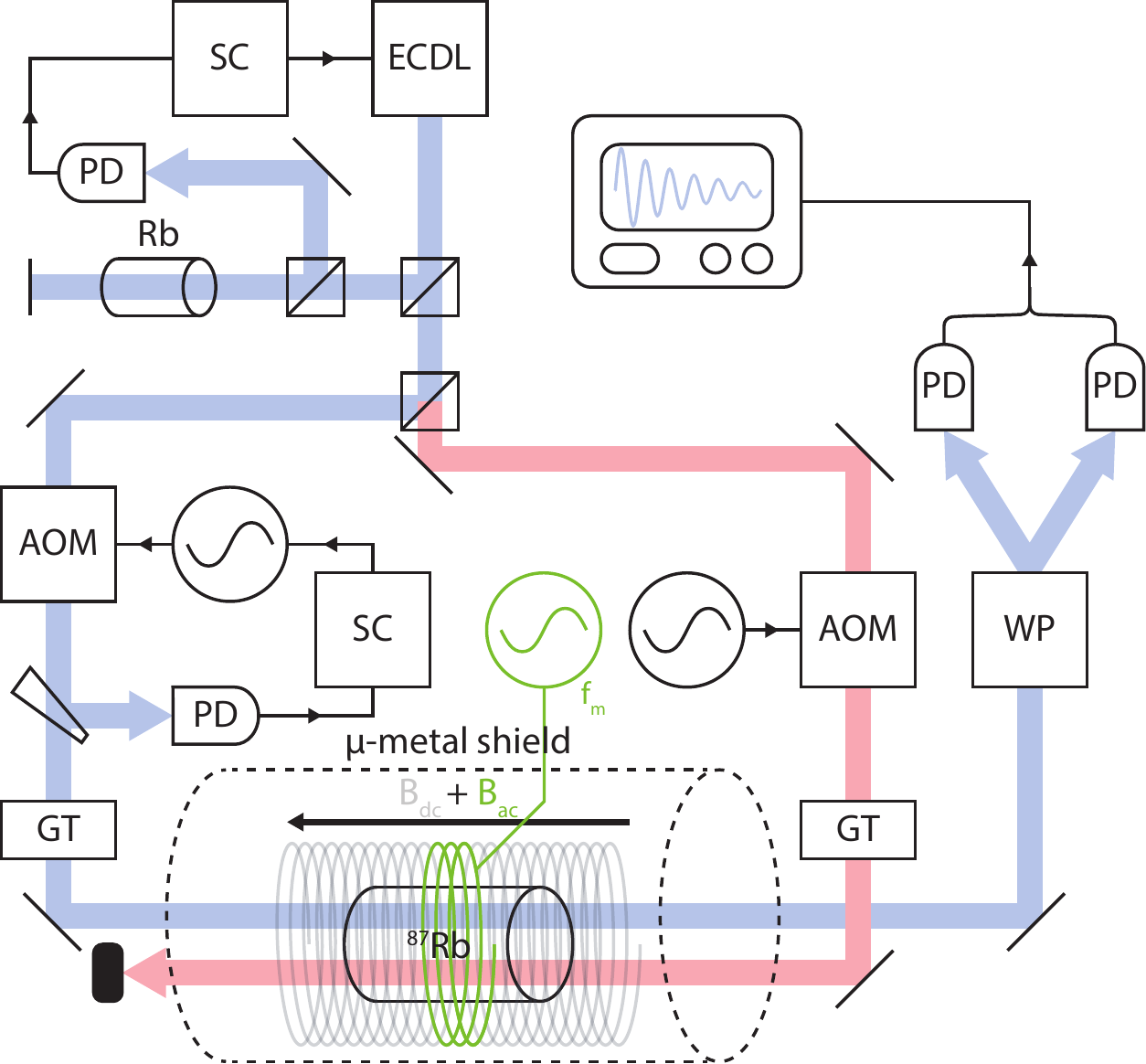}
\caption{\label{fig:setup} Simplified experimental setup, showing the external cavity diode laser (ECDL), acousto-optic modulator (AOM), Glan-Thompson prism (GT), Wollaston prism (WP), photodetector (PD), and servo controller (SC). The probe beam is shown in blue, while the pump beam is shown in red. The modulation coil, shown in green, is wrapped around the center of the solenoid (gray).}
\end{figure}

The experimental setup is presented in Fig.~\ref{fig:setup}. Isotopically pure $^{87}$Rb is contained in a cylindrical vapor cell with a 40\,mm diameter and 40\,mm length. The walls are anti-relaxation coated to extend the coherences between ground-state Zeeman sublevels, with a measured transverse spin-relaxation time of $T_{2} = 45\,$ms (cf. Fig.~\ref{fig:500Hz_RingDown}). The cell remains at room temperature and is housed within a three-layer cylindrical $\upmu$-metal magnetic shield that has a measured shielding factor of approximately $2 \times 10^{3}$. A constant-bias magnetic field of $B_{\text{dc}} \approx \unit[2]{\upmu T}$ is generated along the longitudinal axis of the cell using a solenoid installed within the innermost shield. An oscillatory component of the magnetic field is generated using a separate high-bandwidth coil wrapped around the center of the solenoid.

The atomic vapor is optically pumped and probed using light from an external cavity diode laser tuned near 795\,nm, 80\,MHz below the $5^{2}\textrm{S}_{1/2} \rightarrow 5^{2}\textrm{P}_{1/2}$ (D$_{1}$) transition of $^{87}$Rb. The laser is frequency-locked to the $F = 2 \rightarrow F' = 1$ hyperfine transition ($\gamma/2\pi = 6.9958$\,GHz/T~\cite{Steck}) using saturated absorption spectroscopy in a separate reference cell. The optical pumping beam is amplitude modulated with a 20\% duty-cycle square wave via an acousto-optic modulator at a frequency near $2 f_{\text{L}}$, with a time-averaged power of $\overline{P}_{\text{pump}} \approx 25\,\upmu$W.

The pump and probe beams are linearly polarized by Glan-Thompson (GT) prisms immediately prior to entering opposite sides of the vapor cell, and propagate (anti-)parallel to the magnetic field.
Both beams have a $1/e^{2}$ diameter of 1.5\,mm, and are horizontally displaced from each other by approximately 10\,mm.

After traversing the vapor cell, the probe beam ($P_{\text{probe}} = 5\,\upmu$W) passes through a Wollaston prism, which separates the beam into orthogonal linear-polarization components. These orthogonal components are measured on separate photodetectors, forming a balanced polarimeter. The optical power on each photodetector can be converted into a polarization-rotation angle $\phi(t)$, via~\cite{budker2013optical}:
\begin{equation}
	\phi\left(t\right) = \frac{1}{2}\arcsin\left(\frac{P_{1} - P_{2}}{P_{1} + P_{2}} \right) \, ,
\end{equation}
where $P_{1}$ and $P_{2}$ are the optical powers on the two photodetectors.


\section{Sinusoidal modulation}
\label{sec:SinusoidalModulation}

\begin{figure}[t]
\includegraphics[width=\columnwidth]{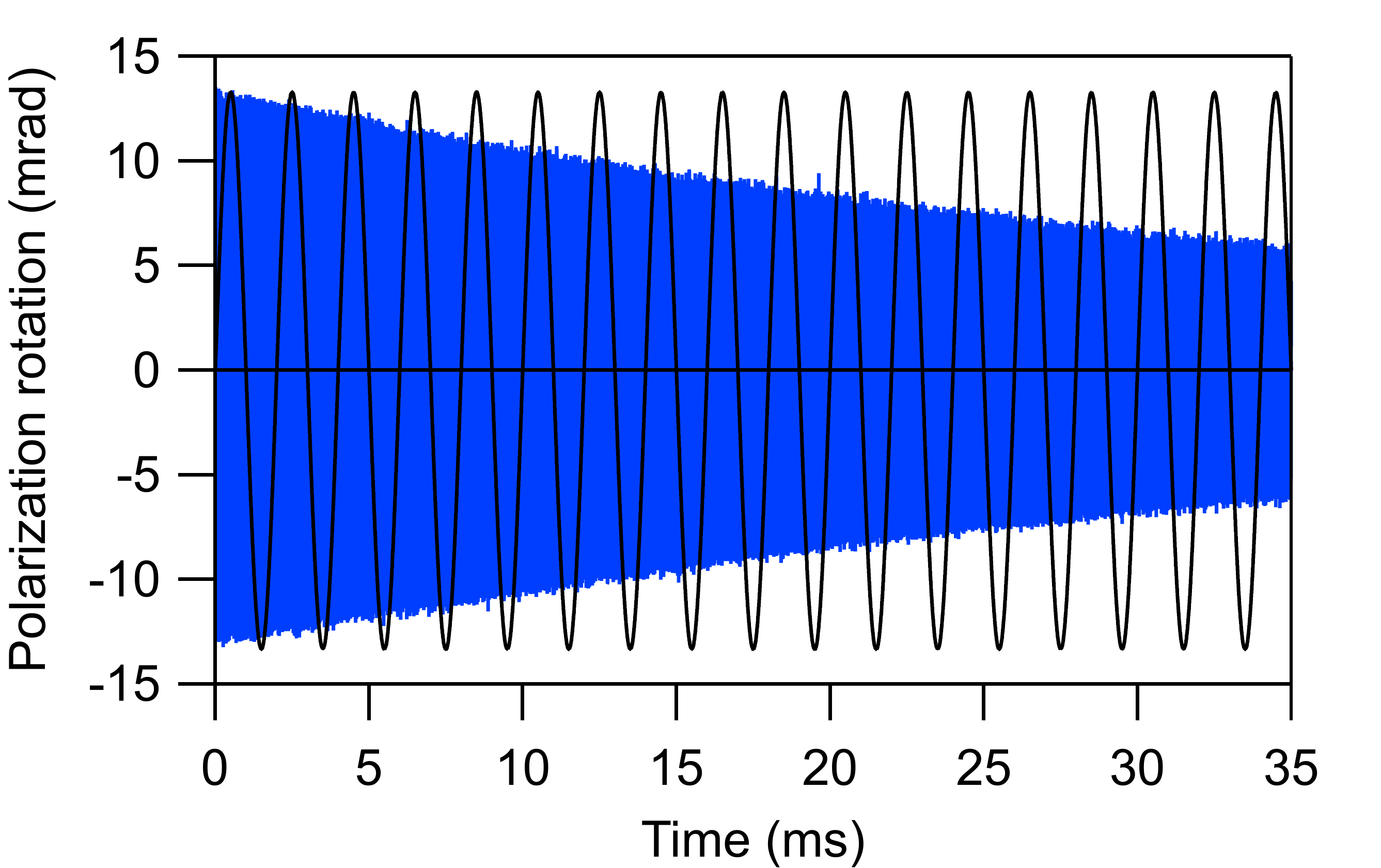}
\caption{\label{fig:500Hz_RingDown} Polarization rotation $\phi(t)$ (blue trace) during free-induction decay of an atomic vapor, measured in a single shot. During the free evolution of the atomic spins, a transient, oscillating magnetic field with $f_{\text{m}} = 500$\,Hz was applied (black trace).}
\end{figure}

\begin{figure}[t]
\includegraphics[width=\columnwidth]{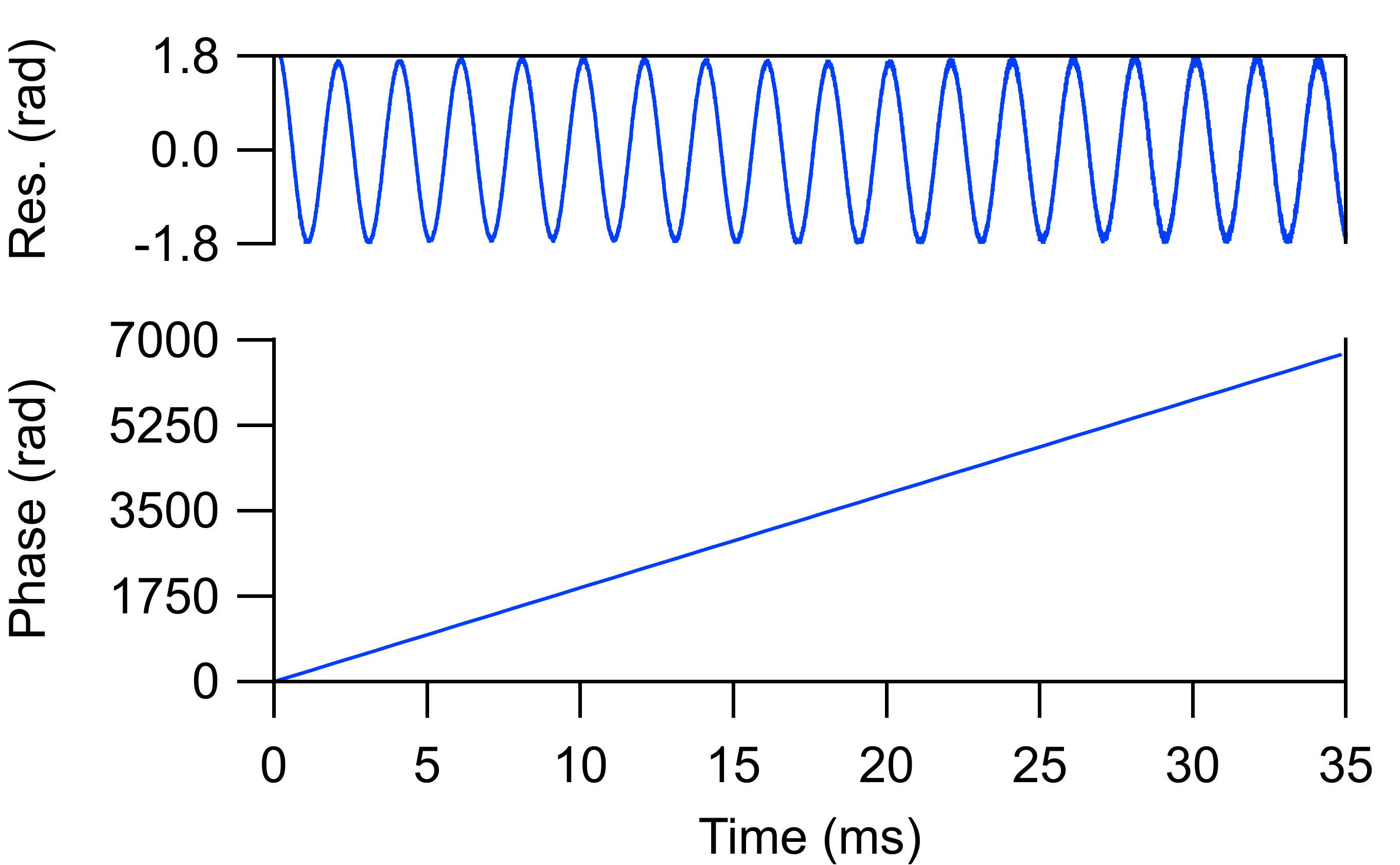}
\caption{\label{fig:500Hz_Modulation} Instantaneous phase $\varphi_{\rm I}(t)$ retrieved from the free-induction decay in Fig.~\ref{fig:500Hz_RingDown}. Linear regression to $\varphi_{\rm I}(t)$ yields a gradient $\text{d}\varphi_{\text{I}}/\text{d}t = 192.49$\,rad/ms, corresponding to a carrier frequency $f_{\rm c} = 2 f_{\text{L}} = 30.636$\,kHz, which implies a dc magnetic-field strength $B_{\text{dc}} = 2.1896$\,$\upmu$T. The residual from the linear fit (top) shows the effect of the sinusoidal magnetic modulation, from which we derive the modulation frequency $f_{\text{m}} = 500$\,Hz, and modulation index $\beta = 1.73$\,rad ($\Delta f = 865$\,Hz) resulting from an ac magnetic-field strength $\Delta B = 61.8$\,nT.}
\end{figure}

Once the transverse spin coherence has been established, the pump beam is extinguished. During the FID, we modulate the magnetic field sinusoidally while measuring the resulting polarization rotation. We retrieve the instantaneous phase of that signal and---per Sec.~\ref{sec:InstantaneousPhase}---are thus able to determine the average magnetic-field strength, as well as the frequency, amplitude, and phase of the magnetic-field modulation. An example of the recorded polarization rotation signal is presented in Fig.~\ref{fig:500Hz_RingDown}, with the process leading to the corresponding instantaneous phase retrieval shown in Fig.~\ref{fig:500Hz_Modulation}. The average gradient of the instantaneous phase can be used to calculate the carrier frequency of the signal using Eq.~\eqref{InstantaneousFrequency}, thus delivering the Larmor frequency and hence the dc magnetic-field strength. The oscillating component of the instantaneous phase has an amplitude $\beta$, proportional to the ac magnetic-field amplitude.

\begin{figure}[t]
\includegraphics[width=\columnwidth]{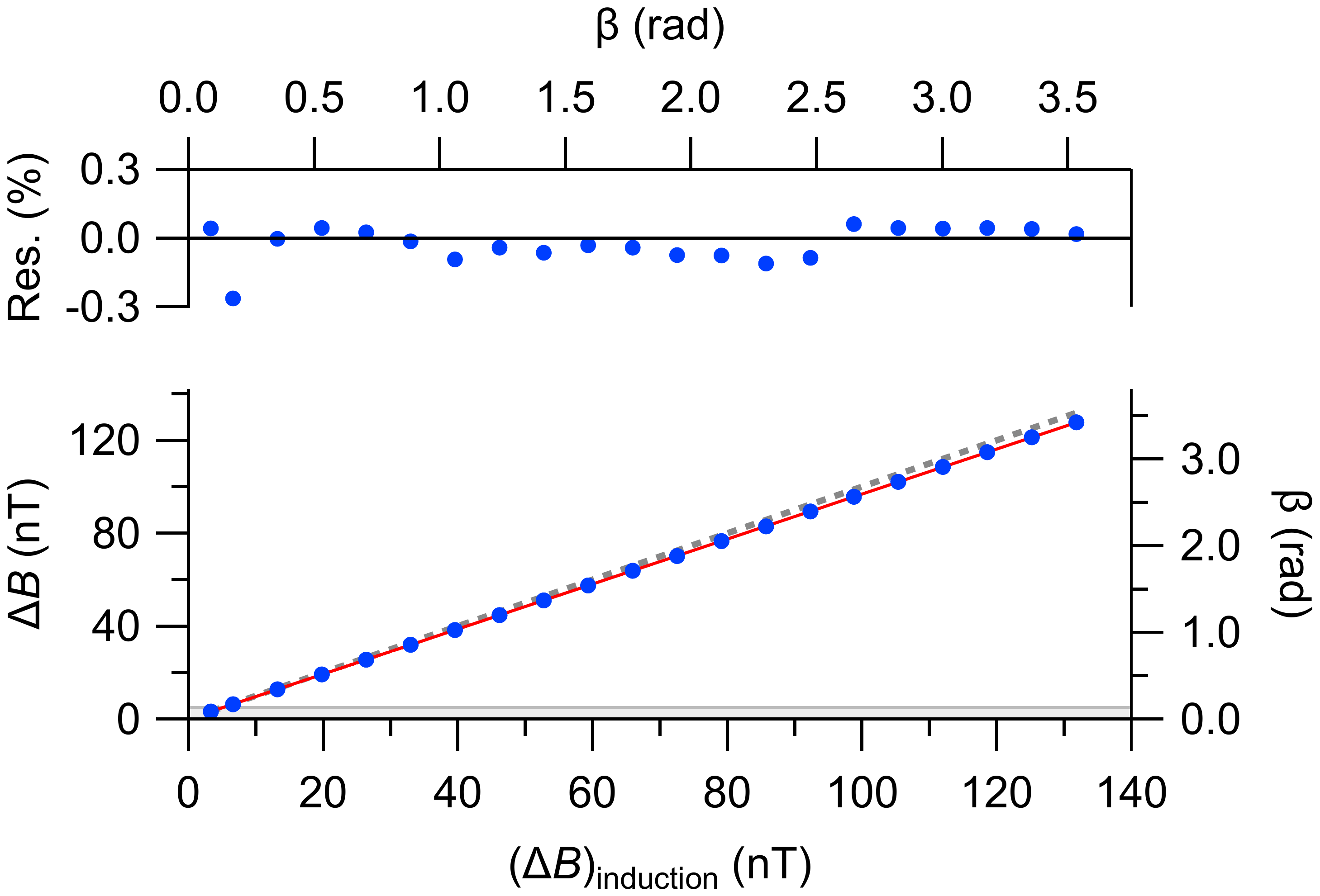}
\caption{\label{fig:522Hz_Linearity} Linear amplitude response of the magnetometer to a sinusoidal modulation at a frequency of $f_{\text{m}} = 522$\,Hz---measured amplitude $\Delta B$ compared with an independent measurement $(\Delta B)_{\text{induction}}$ using a fluxgate magnetometer.
Bottom: Linear-regression (red, solid line) to the data (blue) with a fixed intercept of zero yielded a gradient of 0.9684(1), i.e. a $\sim$$3\%$ deviation from the 1:1 line (gray, dashed). Single standard deviation error bars are indiscernible from the data points here.
Top: Fractional residuals from linear-regression to the data with a fixed, zero intercept.
Shaded in gray is the typical linear amplitude response of conventional radio-frequency magnetometers.
The maximum permissible modulation index under these conditions is $\beta_{\text{max}} \approx 57$\,rad (cf. App.~\ref{sec:MaxModulationIndex}).}
\end{figure}
Our protocol shows an outstanding amplitude linearity and dynamic range to transient fields when compared to traditional rf atomic magnetometers.
In Fig.~\ref{fig:522Hz_Linearity} we present the response of the new device to ac magnetic fields of different amplitudes at a fixed modulation frequency of $f_{\text{m}} = 522$\,Hz. The measured field amplitude $\Delta B$ across a $\sim \unit[140]{nT}$ range was within about 3\% of the value that was independently measured using a fluxgate magnetometer. This discrepancy can be accounted for by the slight difference in measurement volumes between the fluxgate magnetometer and the vapor cell, resulting in a different volume-averaged magnetic-field strength. Linear regression to the data in Fig.~\ref{fig:522Hz_Linearity}, with a fixed offset of zero, yields a gradient of 0.9684(1), corresponding to a linearity of 126\,ppm.
The enhanced linearity and wide, calibration-free amplitude response augurs well for range tracking remote sensing targets of known characteristic size.
A detailed estimation of the maximum permissible oscillating field amplitude is described in App.~\ref{sec:MaxModulationIndex}.

By retrieving the instantaneous Larmor phase, it is possible to detect, in real-time, magnetic field oscillations that are much faster than the Larmor frequency itself. Furthermore, the bandwidth of the sensor is also larger than the Larmor frequency. To demonstrate how our protocol allows access to this experimentally unprecedented regime we detect $f_{\rm m} \leq 50 f_{\rm L}$, over three orders of magnitude greater than the typical bandwidth of rf atomic magnetometers.

\begin{figure}[t]
\includegraphics[width=\columnwidth]{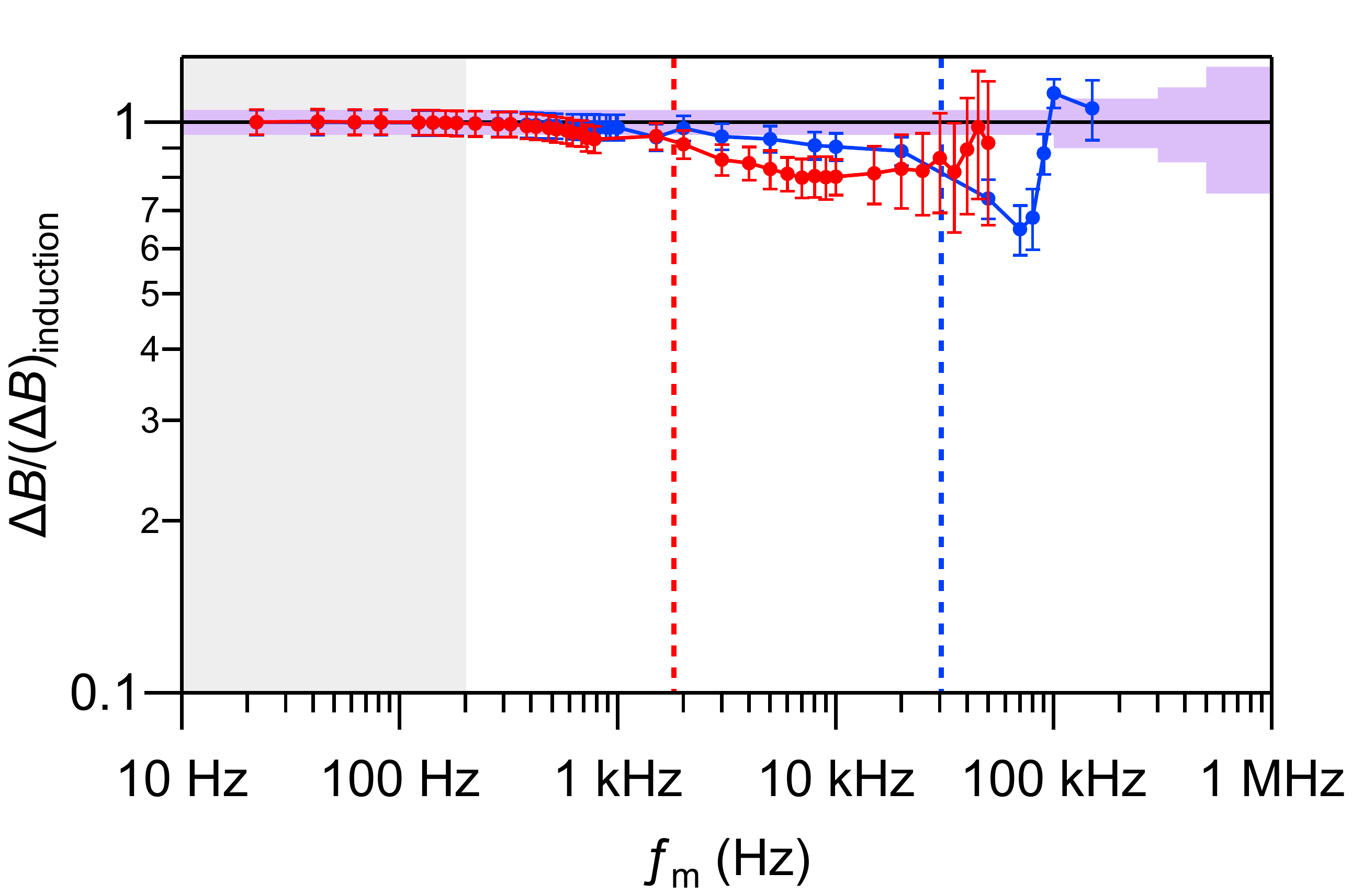}
\caption{\label{fig:Bandwidth} Comparison of the ac field amplitude measured using instantaneous-phase retrieval to an independent inductive measurement versus modulation frequency.
The ratio $\Delta B / (\Delta B)_{\text{induction}}$ is shown for two different carrier frequencies (dashed, vertical lines), $f_{\rm c} = 2f_{\text{L}} = 1.8$\,kHz (red) and $30.5$\,kHz (blue).
Error bars correspond to the combined fractional uncertainty of the induction sensors (specified calibration accuracy of $5$--$25\%$, mauve) and the measurement uncertainty in $\Delta B$ (one standard error).
The gray shaded region denotes the typical frequency response of conventional rf OAMs.
Each datum was measured in a single shot, though averaging permits resolution of higher modulation frequencies, e.g. we observe up to $f_{\text{m}} = 400$\,kHz for 20 shots.
For these data, the modulation index spanned the range $\beta \in \left[0.003, 92\right]$\,rad.
}
\end{figure}

We characterized the frequency response and accuracy of the protocol by comparing the oscillating field amplitude $\Delta B$ imputed using our new instantaneous-phase retrieval to the value $(\Delta B)_{\rm induction}$ determined from either a fluxgate ($f_{\rm m} \leq \unit[1]{kHz}$) or an induction-coil sensor ($f_{\rm m} > \unit[1]{kHz}$).
The ratio of these independent measurements across four decades of modulation frequency is shown in Fig.~\ref{fig:Bandwidth} for two different Larmor/carrier frequencies, corresponding to dc fields $\unit[2.17]{\upmu T}$ and $\unit[128]{nT}$.
This cross-calibration of our FID magnetometer against inductive measurements demonstrates the ability to detect fields modulated well above the Larmor frequency using instantaneous phase-retrieval.
We attribute the small deviations from unity ($\sim \unit[2]{dB}$) in the high frequency parts ($f_{\rm m} > f_{\rm c}$) of the cross-calibration to mutual inductance between the modulation- and induction-coils, rather than any errant atomic response.


\section{Non-trivial modulation}
\label{sec:NonTrivialModulation}

\begin{figure}[t]
\includegraphics[width=\columnwidth]{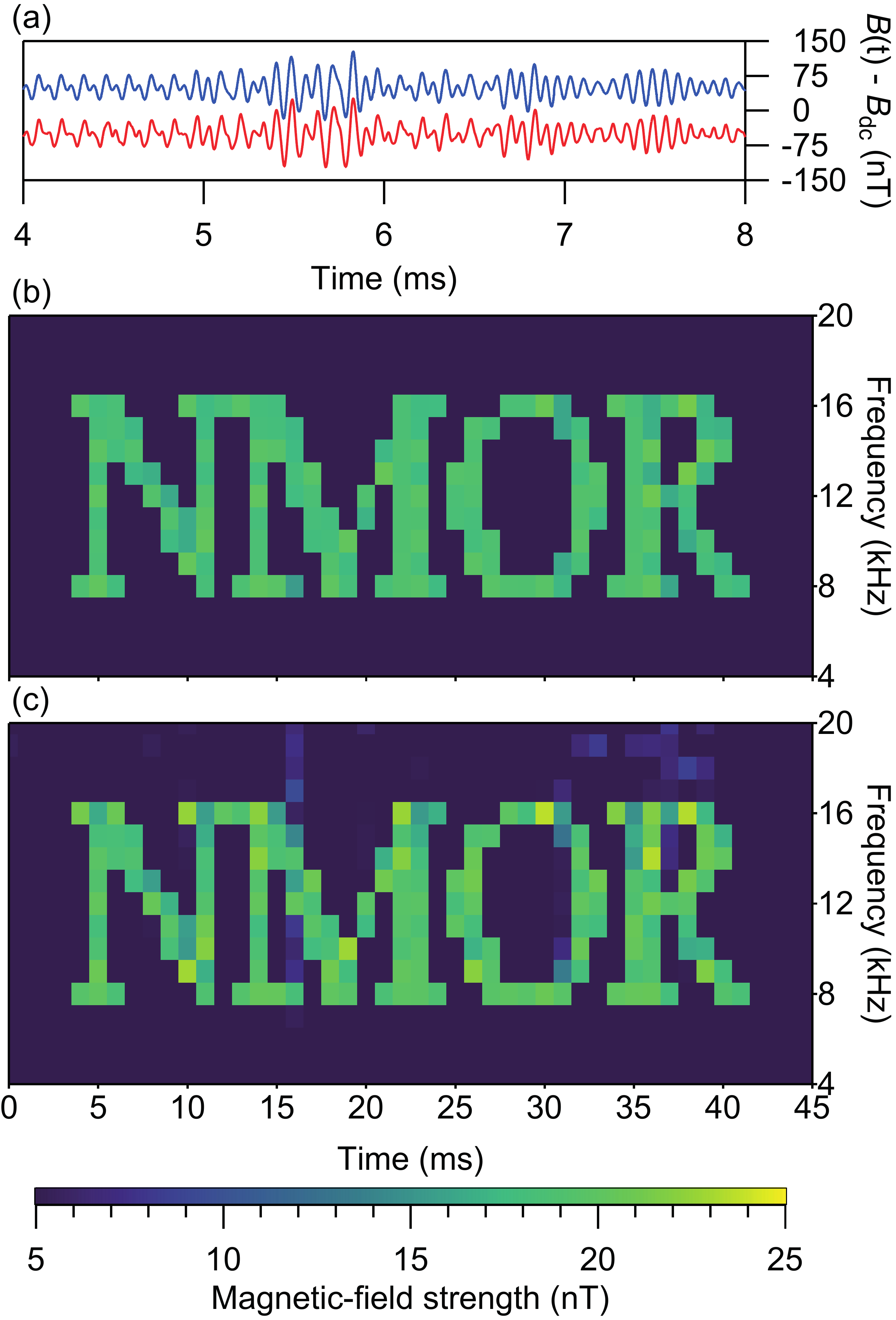}
\caption{\label{fig:Spectrogram} Magnetometer response to a spectrally broad and dynamical magnetic-field modulation. (a) Real-time magnetic field measured using instantaneous phase retrieval (red) compared to the field predicted from the current used to drive the modulation coil (blue). Traces are vertically offset by $\pm \unit[50]{nT}$ for clarity. Spectrograms of this prediction (b) and the measured magnetometer output (c) show the dynamical spectral components ($\Delta B_j$ in Eq.~\ref{NontrivialBfield}.) in close agreement.}
\end{figure}

To highlight the wide bandwidth and large dynamic range of the technique we synthesized a magnetic field with a broad and dynamical spectrum that encodes the acronym `NMOR' (nonlinear magneto-optical resonance).
We drove the modulation coil with a wideband modulated current (see Fig.~\ref{fig:Spectrogram}a), proportional to the resulting time-dependent field:
\begin{equation} \label{NontrivialBfield}
	B(t) = B_{\rm dc} + \sum_{j=1}^{N} \Delta B_{j}(t) \cos\left(2\pi f_{j}t + \varphi_{j}\right) \, .
\end{equation}
Here, we used the same dc magnetic field as displayed in Fig.~\ref{fig:500Hz_Modulation} and $N = 9$ Fourier components, each spaced by $f_{j+1} - f_{j} = 1$\,kHz, where $f_{j} \in \left[8, 16 \right]$\,kHz.
The field amplitudes $\Delta B_{j}(t)$ were piecewise constant, varying abruptly at \unit[1]{ms} intervals to encode the acronym in the time-frequency domain.
The resulting polarization rotation was of the form:
\begin{equation} \label{OptRotSumSinusoidalModulation}
	\phi\left(t\right) = a_{\text{c}}\cos\left(2\pi f_{\text{c}}t + \sum_{j=1}^{N}\beta_{j}(t) \sin\left(2\pi f_{j}t + \varphi_{j}\right) + \varphi_{\text{c}}\right) \, .
\end{equation}
We retrieved the instantaneous phase from this measurand (Sec.~\ref{sec:InstantaneousPhase}), and subtracted a linear fit (Sec.~\ref{sec:SinusoidalModulation}) to impute $\varphi_{\rm I}(t) - 2 \pi f_{\rm c} t - \varphi_{\rm c}$.
The resulting oscillatory component of the instantaneous phase was analyzed in the time-frequency domain using the short-time Fourier transform (STFT).
When the frequency binning is chosen to match $f_{j+1} - f_{j}$, the STFT amplitudes are the time-dependent modulation indices $\beta_j(t) \in \left[11.7, 36.2\right]$\,mrad.
The STFT amplitude was converted to magnetic field amplitudes $\Delta B_j = \pi \beta_j f_j / \gamma$ and is displayed as a spectrogram on Fig.~\ref{fig:Spectrogram}c. We compare this to the spectrogram of the intentionally applied field we predict at the sensor (Fig.~\ref{fig:Spectrogram}b), based on the measured electrical current in the modulation coil and the Biot-Savart law.
The two plots are in excellent agreement with only a small degradation in signal-to-noise ratio of the atomic magnetometer as $t$ approaches $T_{2} = 45$\,ms.


\section{Discussion}

Under the experimental conditions detailed here, the instantaneous phase of the polarization rotation has an observed noise floor of $\unit[68]{\upmu rad_{rms}/\sqrt{Hz}}$ at frequencies above \unit[100]{Hz} --- within 60\% of the shot-noise limit of $\unit[44]{\upmu rad_{rms}/\sqrt{Hz}}$ (cf. App. \ref{sec:NoisePerformance}). This level of phase noise corresponds to a magnetic-field noise of approximately $5 (f / \text{Hz})\,$fT$/\sqrt{\text{Hz}}$. All estimates of the modulation amplitude $\Delta B$, modulation frequency $f_{\text{m}}$, and carrier frequency $f_{\text{c}}$ presented here are within 15\% of the Cram\'{e}r-Rao lower bounds associated with this level of white phase noise~\cite{kay1993fundamentals}.

For modulation frequencies below the carrier ($f_{\text{m}} < f_{\text{c}}$), there is a single frequency component present in the retrieved instantaneous phase, at $f_{\text{m}}$. However when $f_{\text{m}} > f_{\text{c}}$ (supra-carrier modulation), the oscillating component of the instantaneous phase retrieved using Eq.~\eqref{InstantaneousPhaseArg} contains two tones (cf. App.~\ref{sec:InstantaneousPhaseAboveCarrier}), at $f_{\text{m}}$ and $2f_{\text{c}} - f_{\text{m}}$, with equal amplitudes $\beta/2$. The frequency and amplitude of ac magnetic fields can still be imputed in the supra-carrier regime, provided we have knowledge of $f_{\rm c}$. This is guaranteed by the self-certifying estimate of $f_{\rm c}$, which is naturally obtained during linear regression to $\varphi_{\rm I}(t)$ (Fig.~\ref{fig:500Hz_Modulation}). This is equally valid whether $f_{\text{m}} > f_{\text{c}}$ or $f_{\text{m}} < f_{\text{c}}$.

The ability to measure high-frequency magnetic-field fluctuations will, in practice, depend upon the signal-to-noise ratio of the measurement. This can be thought of in two equivalent ways: the measured instantaneous-phase noise is white (independent of frequency), however the signal amplitude, when measured in rotation of the polarization, is given by the modulation index $\beta = \Delta f/f_{\text{m}}$, which scales as $f_{\text{m}}^{-1}$. This results in a magnetic signal-to-noise ratio which also scales as $f_{\text{m}}^{-1}$. Alternatively, if we consider the signal amplitude in terms of the instantaneous-frequency then it is a constant value of $\Delta f$; however, the white instantaneous phase-noise considered in terms of frequency will be violet (differentiated white noise). Once again we see that the signal-to-noise ratio of an ac magnetic field measurement scales as $f_{\text{m}}^{-1}$.

The low-frequency response of this technique is limited to $f_{\text{m}} \gtrsim T_{2}^{-1}$, below which the polarization-rotation signal will decay before a full modulation cycle has been observed. Of course, it would be possible to augment our technique with the traditional approach in which one uses repeated FID measurements: in this case, slow changes in $f_{\rm L}$ can be tracked~\cite{hunter2018waveform}.

We note that typical rf atomic magnetometers detect weak oscillating fields oriented transverse to the static background field~\cite{savukov2005tunable, lee2006subfemtotesla, ledbetter2007detection, chalupczak2012room, savukov2014ultra, deans2018subpicotesla}, whereas our magnetometer senses longitudinally oscillating fields. For arbitrarily oriented rf fields there will be a dead band of $\Delta f_\perp = \gamma \Delta B_\perp / 2\pi$ about the Larmor frequency, where a transverse component of the oscillating field with amplitude $\Delta B_\perp$ can drive Zeeman transitions, resulting in amplitude modulation of the polarization rotation~\cite{anderson2018continuously}.


\section{Conclusion}

We have developed a phase-retrieval technique that can extend the accuracy and applicability of dc optical atomic magnetometers so that they are now suitable for measuring ac magnetic fields. We explore a regime that was previously experimentally inaccessible, where the bandwidth of the sensor exceeds that of the Larmor frequency itself. We have demonstrated calibration-free measurement of oscillating fields in real time with amplitudes up to 150\,nT, and frequencies up to 400\,kHz, an increase in the amplitude and frequency response of two and three orders of magnitude over conventional rf atomic magnetometers respectively. The instantaneous-phase retrieval can be applied more broadly to quantum sensors that employ weak continuous measurement of spin precession, including those exploiting synchronous detection and feedback~\cite{vijay2012stabilizing}. This capability augurs well for robust field-deployable magnetometers which, unlike lab-based apparatus, face inherently unpredictable operational environments and transient signals of interest.


\section*{Acknowledgments}

The authors acknowledge financial support from the Defence Science and Technology Group, and the South Australian Government through the Premier's Science and Research Fund. N. W. and P. L. would like to thank Yvonne Stokes for useful discussions.

\appendix

\section{Maximum Permissible Modulation Index}
\label{sec:MaxModulationIndex}

The frequency-modulated polarization rotation deriving from an oscillating magnetic field (Eq.~\ref{OptRotSinusoidalModulation}) can be rewritten using the Jacobi-Anger expansion~\cite{korsch2006two}:
\begin{equation} \label{OptRotBessel}
	\phi\left(t\right) = a_{\text{c}}\sum_{n = -\infty}^{\infty}J_{n}\left(\beta\right)\cos\left(2\pi\left[f_{\text{c}} + nf_{\text{m}}\right]t + \varphi_{\text{c}} + n\varphi_{\text{m}}\right) \, ,
\end{equation}
where $J_{n}$ is the $n^{\text{th}}$-order Bessel function of the first kind. Thus, the polarization rotation comprises an infinite series of sidebands, distributed symmetrically about the carrier at $f_c$, spaced by integer-multiples of the modulation frequency $f_{\text{m}}$, with amplitudes $a_{n} = a_{\text{c}}J_{n}\left(\beta\right)$.

\begin{figure}[t]
\includegraphics[width=\columnwidth]{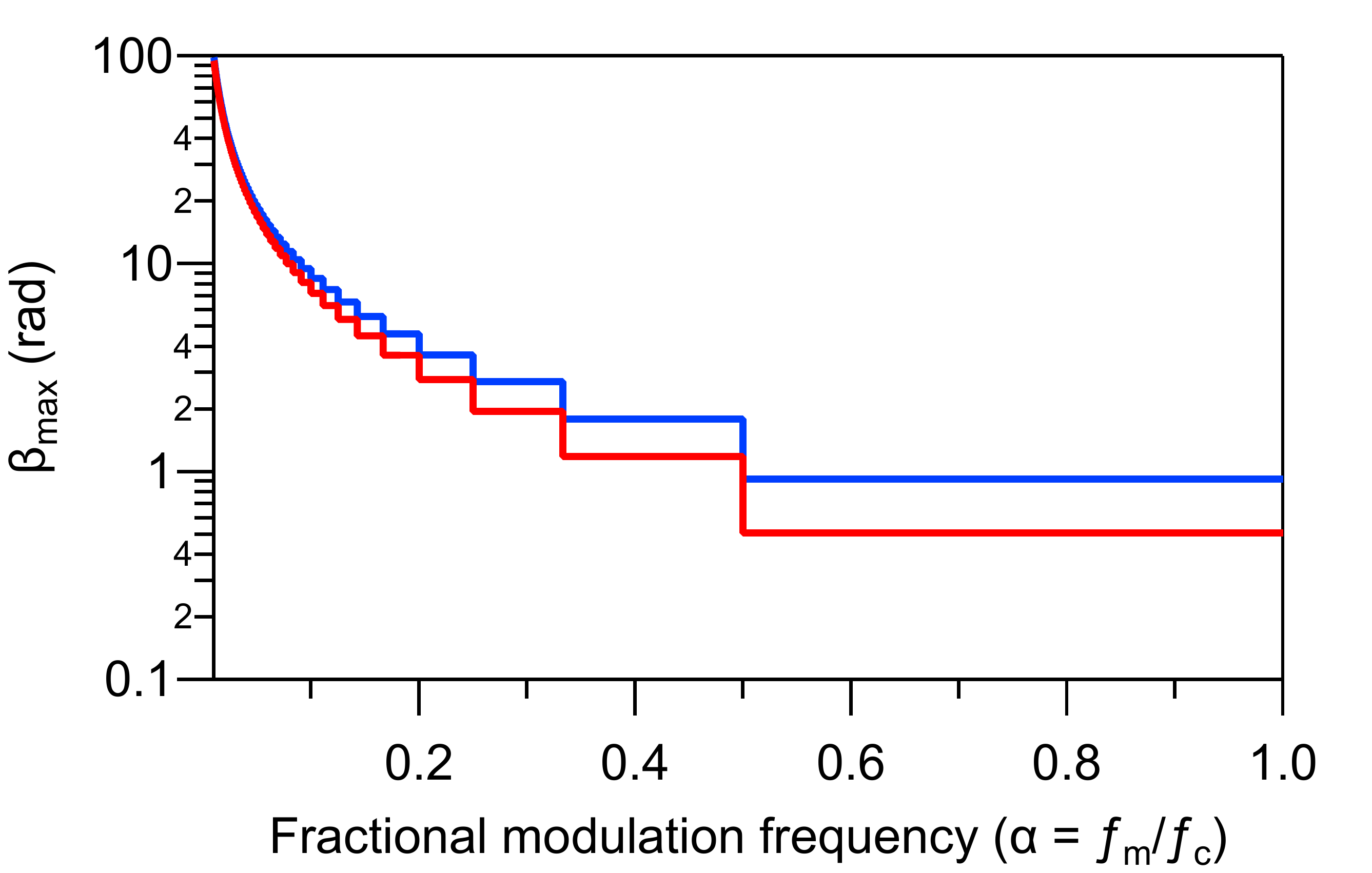}
\caption{\label{fig:BetaMax} Maximum permissible modulation index, $\beta_{\text{max}}$, calculated using Eq.~\eqref{BesselUpperLimit2} for $P_{\text{max}} = 0.01$ (blue trace) and $P_{\text{max}} = 0.001$ (red trace). For $\alpha > 1$, $\beta_{\text{max}}$ remains constant until $\alpha \approx (4 P_{\rm max})^{-1/2}$, beyond which the constraint that $B(t) > 0$ dominates, and then $\beta_{\text{max}} = \alpha^{-1}$.}
\end{figure}

A conservative upper bound for the maximum permissible modulation index, $\beta_{\text{max}}$, is derived by considering the fractional power in sidebands about $f_{\rm c}$ which extend to negative frequencies. These sidebands are discarded when numerically computing the analytic signal, which can compromise the instantaneous-phase retrieval. For a given fractional modulation frequency, $\alpha = f_{\text{m}}/f_{\text{c}}$, the largest value of $n$ such that $f_{\text{c}} - nf_{\text{m}} > 0$, is given by:
\begin{equation}
	n_{\text{max}} = \text{ceil}\left(\alpha^{-1}\right) - 1 \, .
\end{equation}
To calculate $\beta_{\text{max}}$ for a given $\alpha$, one must choose an acceptable fractional power $P_{\text{max}}$ contained in the discarded sidebands. This constraint amounts to:
\begin{equation} \label{BesselUpperLimit1}
	\sum_{n = n_{\text{max}} + 1}^{\infty} J_{n}\left(\beta_{\text{max}}\right)^{2} \leq P_{\text{max}} \, .
\end{equation}
Using Bessel-function identities and symmetry relations, Eq.~\eqref{BesselUpperLimit1} simplifies to:
\begin{equation} \label{BesselUpperLimit2}
	\sum_{n = -n_{\text{max}}}^{n_{\text{max}}} J_{n}\left(\beta_{\text{max}}\right)^{2} \geq 1 - 2P_{\text{max}} \, .
\end{equation}
The maximum permissible modulation index calculated using Eq.~\eqref{BesselUpperLimit2} is shown in Fig.~\ref{fig:BetaMax}, for two different power fractions: $P_{\text{max}} = 0.01$ and $0.001$. Due to the nonlinearity of Eq.~\eqref{BesselUpperLimit2}, an increasingly stringent bound on $P_{\text{max}}$ does not significantly reduce $\beta_{\text{max}}$.

An additional constraint on $\beta$ arises from requiring that the instantaneous magnetic field never be zero, which could result in non-adiabatic spin-flips that compromise this measurement.
This limits the maximum measurable ac magnetic-field amplitude to $\Delta B < B_{\rm dc}$, equivalent to $\Delta f < f_{\rm c}$ and thus $\beta < f_{\rm c} / f_{\rm m} = \alpha^{-1}$.
This constraint becomes stricter than the sideband-power considerations for $\alpha \gtrsim (4 P_{\rm max})^{-1/2}$.

\section{Instantaneous Phase Estimation when Modulating Faster Than the Carrier Frequency}
\label{sec:InstantaneousPhaseAboveCarrier}

For supra-carrier modulation ($f_{\text{m}} > f_{\text{c}}$), the lower sideband nearest the positive-frequency carrier ($f = f_{\rm c}$) at $f = f_{\text{c}} - f_{\text{m}}$ becomes negative ($n=-1$ in Eq.~\ref{OptRotBessel}). When numerically computing the analytic signal in this regime, this sideband is discarded. However, due to the Hermitian symmetry of the Fourier transform of a real function, there is a corresponding upper sideband of the negative-frequency carrier ($f = -f_{\rm c}$) at $f = -f_{\text{c}} + f_{\text{m}}$ which becomes positive. In the case of weak modulation (i.e. $\beta \lesssim 1$, which is generally the case when $f_{\text{m}} > f_{\text{c}}$), the majority of the modulated power of the analytic signal is contained in the two positive frequencies $f = f_{\text{c}} + f_{\text{m}}$ and $f = -f_{\text{c}} + f_{\text{m}}$. The spacing of these tones from the positive-frequency carrier $f_{\text{c}}$ is $f_{\text{m}}$ and $\abs{2f_{\text{c}} - f_{\text{m}}}$, respectively. The asymmetry of these dominant sidebands results in a modification of the instantaneous phase retrieved using Eq.~\eqref{InstantaneousPhaseArg}, in the form of double-sideband suppressed-carrier amplitude modulation:
\begin{widetext}
\begin{equation}
\begin{aligned}
\varphi_{\text{I}}\left(t\right) & = 2\pi f_{\text{c}}t + \frac{\beta}{2}\sin\left(2\pi f_{\text{m}}t + \varphi_{\text{m}}\right) + 
 \frac{\beta}{2}\sin\left(2\pi \left(2f_{\text{c}} - f_{\text{m}} \right)t - \varphi_{\text{m}}\right) + \varphi_{\text{c}} \\
 &= 2\pi f_{\rm c} t + 
 \beta \cos\left( 2\pi(f_{\text{m}} - f_{\text{c}}) t + \varphi_{\text{m}} \right) 
 \sin\left( 2\pi f_{\text{c}} t \right) + \varphi_{\text{c}} \, .
\end{aligned}
\end{equation}
\end{widetext}
The resulting spectrum of $\varphi_{\text{I}}\left(t\right) - 2\pi f_{\text{c}}t - \varphi_{\rm c}$ contains two tones, at $f_{\text{m}}$ and $2f_{\text{c}} - f_{\text{m}}$. However, as the power is evenly distributed between these two tones, and the method prescribed in Fig.~\ref{fig:500Hz_Modulation} can still be used to determine $f_{\rm c}$, the frequency and amplitude of ac magnetic fields can still be imputed unambiguously in the supra-carrier regime.

\section{Sensitivity}
\label{sec:NoisePerformance}

\begin{figure}[t]
	\includegraphics[width=\columnwidth]{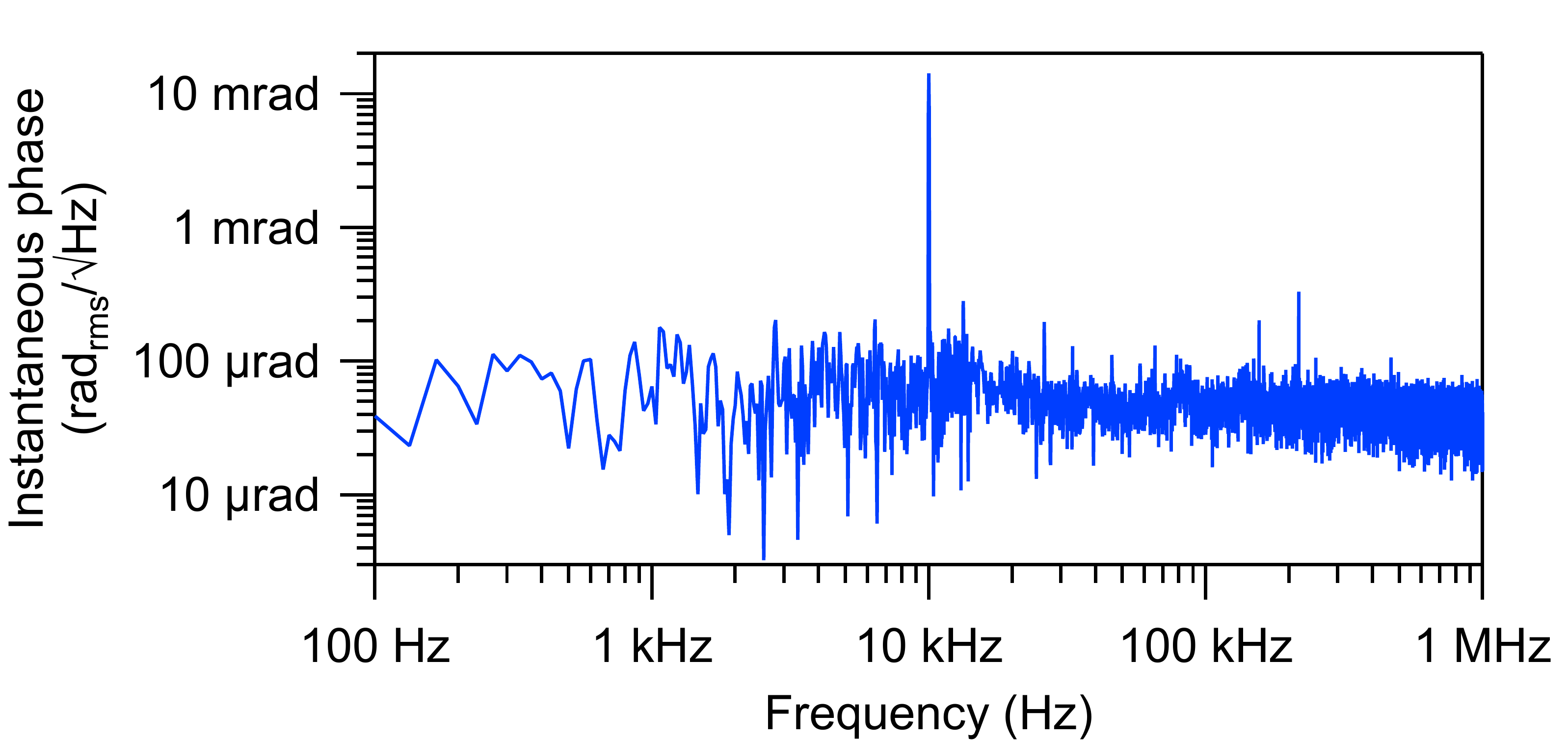}
	\caption{\label{fig:NoisePerformance} Amplitude spectral density of the instantaneous phase for a carrier frequency $f_{\rm c} = 2 f_{\rm L} = \unit[30.636]{kHz}$, and with an applied magnetic-field modulation of amplitude $\Delta B = \unit[102]{nT}$ at a frequency $f_{\rm m} = \unit[10]{kHz}$.}
\end{figure}

The magnetic sensitivity of this technique is ultimately determined by the ability to resolve small changes in the instantaneous phase.
A representative amplitude spectral density of the instantaneous-phase is presented in Fig.~\ref{fig:NoisePerformance}, indicating a nearly white noise floor with magnitude $\unit[68]{\upmu rad_{rms}/\sqrt{Hz}}$.
Using Eqs.~\eqref{InstantaneousField} and \eqref{InstantaneousFrequency}, this level of instantaneous phase noise corresponds to a magnetic-field noise of $\unit[5.0 (f / \text{Hz})]{fT/\sqrt{Hz}}$.

A photocurrent-referenced measurement of the optical power indicates the shot-noise limit for instantaneous-phase is $\unit[44]{\upmu rad_{rms}/\sqrt{Hz}}$, or about $\unit[3.1 (f / \text{Hz})]{fT/\sqrt{Hz}}$ in magnetic units.

%

\end{document}